\documentclass[12pt,preprint]{aastex}

\usepackage{epsfig, amsmath, amsfonts}

\newcommand\bb[1] {   \mbox{\boldmath{$#1$}}  }

\newcommand\del{\bb{\nabla}}

\def\gtsima{$\; \buildrel > \over \sim \;$}
\def\gtsim{\lower.5ex\hbox{\gtsima}}

\newcommand\bcdot{\bb{\cdot}}

\newcommand\dd{\partial}

\newcommand\beq{ \begin{equation} }
\newcommand\eeq{ \end{equation} }
\def\spose#1{\hbox to 0pt{#1\hss}} 
\def\ltsim{\mathrel{\spose{\lower.5ex\hbox{$\mathchar"218$}}
\raise.4ex\hbox{$\mathchar"13C$}}}

\shorttitle{Keplerian Wave Solution}
\shortauthors{Balbus}

\begin{document}

\title{An Exact, Three-Dimensional, Time-Dependent Wave Solution in Local
Keplerian Flow}

\author{Steven A. Balbus}
\affil{\'Ecole Normale Sup\'erieure, Laboratoire de Radioastronomie,
24, rue Lhomond,\\
75231 Paris CEDEX 05, France}
\author {John F. Hawley}
\affil{Astronomy Department, University of Virginia, Charlottesville, VA 22903}



\begin{abstract}
We present an exact three-dimensional wave solution to the shearing sheet
equations of motion.  The existence of this solution argues
against transient amplification as a route to turbulence in unmagnetized
disks.  Moreover, because the solution covers an extensive dynamical range
in wavenumber space, it is an excellent test of the dissipative properties
of numerical codes.  
\end{abstract}


\section{Introduction}

The combination of magnetic fields and differential rotation is a volatile
one, leading to the magnetorotational instability, or MRI, under very
general conditions (e.g. Balbus 2003).  Laboratory evidence for the MRI
in liquid sodium has recently been claimed
(Sisan et al. 2004), 
and there can be little doubt that in magnetized disks the MRI will be
present and a source of the enhanced angular momentum transport needed for
accretion to proceed.  But not all astrophysical disks need be 
everywhere ionized at a level at which the field and the gas are well-coupled.
Therefore the question of whether or not an unmagnetized, laminar gas
in Keplerian rotation is vulnerable to turbulent breakdown continues to
attract attention.

The nature of the local stability of Keplerian disks has been
investigated by detailed numerical simulation (Balbus, Hawley, \& Stone
1996; Hawley, Balbus, \& Winters 1999).  There is no evidence of nonlinear
instability.  Indeed, the notion that Keplerian disks are intrinsically
turbulent would be most surprising and problematic to MHD numericists,
who routinely check the validity of an MHD code by ensuring that the
disk quickly returns to stability when the magnetic field is removed.
Even when nonlinear perturbations are directly seeded, they do not lead
to turbulence.

Numerical simulations have been criticized, however, on the grounds that
they lack the necessary resolution to reveal nonlinear instabilities
(e.g. Richard \& Zahn 1999; Afshordi,  Mukhopadhyay, \& Narayan 2005). 
The claim is that the effective Reynolds number of available codes is
too low, and that the numerical schemes dissipate nascent nonlinear
instabilities.  In some cases, a specific destabilizing mechanism is suggested.
Afshordi et al. (2005) suggest, for example, that the transient growth exhibited by
linear nonaxisymmetric perturbations, in the course of evolving
from leading to trailing wavenumbers, might become sufficiently large to
cross a threshold and initiate nonlinear turbulence.

The first systematic numerical studies of Keplerian stability are
now a decade old, and numerical codes have developed and improved.
Our physical understanding of disks has also deepened.  We are motivated
to reexamine the problem of the stability of Keplerian disks.   In this
paper, we present precise analytic solutions for three-dimensional
incompressible perturbations in Keplerian disks, recovering the previously
known two-dimensional solutions as special cases.  We show that in fact
all such solutions are valid both linearly and nonlinearly because the
nonlinear terms in the equations of motion vanish.  Thus, these solutions
by themselves do not trigger nonlinear instabilities, regardless of
the level of transient amplification attained.  Since the solutions
exhibit both strongly leading and strongly trailing large wavenumbers,
they cover a very large dynamical range.  They are, in fact, a good
test of the dissipation properties of numerical codes.  Conversely,
the numerical codes test the stability of these exact time-dependent
solutions, which would otherwise be a difficult analytic calculation.

Recently, laboratory experiments have weighed in on the controversy.
A high Reynolds number ($\sim 10^6$) Couette flow experiment found
that Keplerian-like profiles are nonlinearly stable (Ji 2006, private
communication).  This finding is compatible with numerical simulations,
but in conflict with an earlier experiment (Richard 2001)
that claimed that such profiles were unstable.  The Ji experiment uses
differentially rotating endcaps to break up Ekman boundaries (which 
otherwise cause unwanted interior circulation), and Doppler velocimetry
techniques to verify that the rotation profiles matched their analytic
counterparts.  The Reynolds stress in the Ji experiment has also been
measured, and residual fluctuations in the radial and azimuthal velocities
are uncorrelated: the tensor vanishes on average. 

These results are compelling reasons to believe that local Keplerian flow
is nonlinearly stable, and that the numerical simulations have not been
prohibitively dissipative or otherwise qualitatively misrepresentative.
The explicit solutions we present in this paper will provide numericists
with a well-posed numerical test bed that will sharpen our understanding
of local small scale behavior in Keplerian disks.

An outline of this paper is as follows.  In \S 2 we provide an overview of
the issues that have been debated.  Section 3 sets up the mathematical
background, and the full analysis is given in \S 4.  The solutions
derived in \S 4 are applied as a code test problem in \S 5.  Finally,
\S 6 summarizes our work.

\section {A Brief Review}

Axisymmetric
waves propagating in the interior of an unmagnetized
disk are stable if and only if the specific angular momentum
$l$ increases with cylindrical radius $R$.
This is the classical Rayleigh stability criterion,
and it is generally satisfied in astrophysical disks.  
When stable, fluid displacements in the disk plane
oscillate about their equilibrium
circular orbits with a characteristic frequency $\kappa$.
The value of $\kappa^2$ is
directly
proportional to the specific angular momentum gradient
$l$.  In particular,
\beq\label{eqn:epicycle}
\kappa^2 = \frac{1}{R^3} \frac{dl^2}{dR} 
= 4\Omega^2 + \frac{d\Omega^2}{d\ln R},
\eeq
where $R$ is the cylindrical radius
and $\Omega =l/R^2$ is the local disk angular velocity.  The formal
transition to a negative value of $\kappa^2$ corresponds to the onset of
rotational instability.  It is to be noted that this result applies only to
axisymmetric disturbances; there is at present no known rigorous
nonaxisymmetric criterion.  (Hence the ongoing debate.)
A radially discontinuous rotation
profile, for example, would be unstable to
non-axisymmetric disturbances by the Kelvin-Helmholtz
instability, even if the rotation rate increased outward
and was formally Rayleigh-stable.

For a Keplerian disk, $\kappa^2=\Omega^2$, which is a stable profile
by the Rayleigh criterion.  Absent pressure forces, this is 
unconditionally stable: the behavior of Keplerian orbits is not an object of
current debate.  Pressure forces must, therefore, be at the heart of any
proposed instability mechanism.  This is already a potential difficulty
for proponents of instability, since any system that is close to Keplerian
must be characterized by very small radial background pressure gradients,
and perturbative pressure gradients generally behave acoustically.
Even {\em global} instabilities in flows that would be
{\em locally} Rayleigh-stable
revert to stable behavior as the disk approaches a Keplerian profile
(see, e.g., Goldreich, Goodman, \& Narayan 1986).

The claims to nonlinear instability discussed in the Introduction are
of course meant to apply to systems which are linearly stable.  It is
not unprecedented for linear waves to be subject to a finite amplitude
instability (e.g., Infeld \& Rowlands 1990), and even at arbitrarily
small amplitudes, one-dimensional adiabatic sound waves show nonlinear
steepening behavior.  (In fact, the shearing wave solutions to be
discussed are, in their tightly-wrapped phase, subject to dissipative
instabilities that may well prevent them from achieving even modest
amplitudes.)  But proponents of nonlinear Keplerian instability are
arguing for something far more novel than wave modification.  The claim
is that well-behaved local linear disturbances that formally begin
and end their existence as small amplitude waves, heavily dominated
by restoring forces, in actuality become self-sustaining turbulence.
If only the small nonlinear terms were properly resolved, it is argued,
the simulations would reveal this behavior.  In this view, current disk
codes would be not merely imprecise, but grossly misrepresentative of
the most fundamental character of Keplerian flow.

Planar Couette and Poiseuille viscous flows are two classic shear profiles
that exhibit a breakdown into turbulence.  Their behavior has motivated
the notion that Keplerian rotation ought to do the same.  Planar Couette
flow is linearly stable (in a marginal sense) but nonlinearly unstable,
whereas Poiseuille flow is linearly unstable if the appropriately defined
Reynolds number $Re$ exceeds a large critical value ($Re_{crit}=5772$,
see e.g., Drazin \& Reid [1981] for details).  Both flows are exquisitely
sensitive to nonlinear disturbances.

Planar Poiseuille and Couette flows have boundary layers and linear
dynamical responses lacking a dominant restoring force.  The
presence of viscous boundary layers means that the Reynolds number is an
important feature of the flow stability, and at the same time allows for
{\em linear} instability in Poiseuille flow at sufficiently high Reynolds
number.  By way of contrast, local Keplerian flow
has no viscous boundary layers, and its destabilizing shear is strongly
stabilized by the Coriolis force, $2\Omega >|d\Omega/d\ln R|$.  It is not
surprising, therefore, that the response of disks is sensitive neither
to the Reynolds number, nor to the amplitude of the initial disturbances.

The mechanism by which shear flow nonlinearly breaks down into turbulence
seems to be fairly-well understood, at least in its essentials: nonlinear
perturbations are in essence linear disturbances that feed off of a
marginally stable, finite amplitude two-dimensional perturbation that is
imposed on the shear flow (Zahn, Toomre, \& Spiegel 1974; Bayly, Orszag,
\& Herbert 1988).  If no such marginally stable, finite amplitude state
exists---and for Keplerian disks it does not---the mechanism breaks down.
In fact, detailed numerical simulations show no tendency for Keplerian
rotation to evolve toward self-sustained turbulence, even when driven
vigorously well into the nonlinear regime by a sympathetic programmer.
A revealing example is the case of thermally-driven convective turbulence.
In the Boussinesq limit, this results in a very small but measurable
angular momentum flux, with the wrong sign to sustain turbulence (Stone \&
Balbus 1996).\footnote{A negative Reynolds stress, which cannot be due to
numerical dissipation, poses difficulties for the notion that Keplerian
disks are formally turbulent, but at a level too small to be detected
(Lesur \& Longaretti 2005).  Shear turbulence, whatever its magnitude,
can be sustained only by a {\em positive} Reynolds stress.} Moreover,
even codes with very different small scale dissipation properties converge
when presented with the same initial set of finite amplitude disturbances
(Hawley et al. 1999).   Such convergence would hardly be expected if
each code was anomalously dissipative in its own way.


The current work is supportive of numerical simulation and the recent
Ji laboratory experiment.  The analytic perturbation solutions are in
fact {\em exact} solutions to the dynamical equations.   The nonlinear
terms that are supposed to trigger turbulence instead vanish.  This is
problematic for the idea that these disturbances seed turbulence.  More
positively, it is an opportunity to test the accuracy of current codes.
The solutions may be determined to any desired order of accuracy either
by well-established WKB techniques, or by direct numerical integration
of a simple ordinary differential equation.  The wave forms involve
time-dependent wavenumbers of arbitrarily high magnitudes, both leading
and trailing.  Despite the fact that the amplitude of the departure
from Keplerian flow can be made arbitrarily high, the solution remains
formally valid.  Secondary instabilities feeding off this finite amplitude
wave cannot be excluded {\em a priori,} but in fact the recent findings of
Shen, Stone, \& Gardiner (2006) suggest that such secondary instabilities
make matters even more difficult.  Very tightly-wound leading wave solutions
(which would evolve toward the largest amplitudes) break down in the
early stages of their evolution by what seems to be a Kelvin-Helmholtz
instability.  Far from becoming nonlinear, the wave dissipates before
there is any appreciable growth.

\section{Preliminaries}
\subsection{The Shearing Box}

The {\em shearing box} is a mathematical limit of the equations
of motion in cylindrical coordinates that corresponds to 
retaining the local behavior of differentially rotating
flow.  The approach has a long history,
dating at least as far back as the classical two-dimensional
{\em shearing sheet} calculation of
Goldreich \& Lynden-Bell (1965).  A more recent MHD description
can be find in the review article of Balbus \& Hawley (1998).
The idea is that one looks at the flow very near a fixed 
cylindrical radius
$R$, and takes the limit that $R\rightarrow\infty$.  The angular
velocity $\Omega$ remains finite, and the azimuthal
velocity $v_\phi$ formally increases without bound. 
We are interested in departures from the Keplerian
profile $v_\phi = v_K$, defined by
\beq
\bb{u} = \bb{v} - v_K\bb{e_\phi},
\eeq
where $\bb{v}$ is the exact velocity, and $\bb{e_\phi}$
is a unit vector in the azimuthal direction.  Note that the $\bb{u}$ velocities
are taken relative to the radially changing Keplerian profile,
not relative to a frame that is rotating at a constant angular
velocity.  The unperturbed flow is thus $\bb{u}=0$, and $|\bb{u}|$
is always small compared with the large quantity $v_\phi$. 

In a standard $R,\phi, z$ cylindrical coordinate system,
the inviscid equations of motion in the local shearing box limit are
(Balbus \& Hawley 1998):
\beq\label{radial}
\frac{Du_R}{Dt} - 2\Omega u_\phi = -\frac{1}{\rho}\frac{\dd P}{\dd R},
\eeq
\beq\label{azi}
{Du_\phi\over Dt} +{\kappa^2\over 2\Omega} u_R = -{1\over \rho R}{\dd P\over
\dd \phi},
\eeq
\beq
{Du_z\over Dt} = -{1\over\rho} {\dd P\over \dd z} - z\Omega^2,
\eeq
where 
\beq
{D\ \over Dt} = {\dd\ \over\dd t} + \Omega {\dd \over \dd\phi}
+\bb{u\cdot}\del.
\eeq
Our notation is as follows: $\rho$ is the mass density of the disk,
$P$ is the gas pressure, and 
$\kappa^2$ is the epicyclic frequency associated
with departures from circular orbits (eqn.~\ref{eqn:epicycle}).
Formally, 
the local vertical gravitational force 
is $-z\Omega^2$ for a Keplerian potential.  However it is often ignored
in shearing box calculations (the ``cylindrical approximation'')
and we shall generally do so in this paper.

As noted, the equations apply to a local radial neighborhood around
a large radius, $R$.  The coefficients $2\Omega$,
$\kappa^2/2\Omega$, and $\Omega^2$ 
are regarded as local constants.
Finally, the equation for mass conservation in our constant
density system is simply
\beq
\del\bcdot\bb{u} = 0 .
\eeq
It is to be emphasized that these equations, while local, are fully nonlinear.

It is also informative to write the shearing box equations 
in a frame that is rotating at a fixed angular velocity.
Fix $R$ to be the radius at the center of the box, and
fix $\Omega=\Omega(R)$.
Let $x$ be the radial distance from $R$, $y=R(\phi-\Omega t)$,
and $z$ the vertical distance from the midplane.  With
\beq
\bb{w} = \bb{v} - R\Omega\bb{e_\phi},
\eeq
in their most general form 
the shearing box dynamical equations are
\beq\label{hill1}
\left( {\dd \over \dd t} + \bb{w\cdot}\del\right){w_R}
- 2\Omega w_\phi = - x{d\Omega^2\over d \ln R} - {1\over\rho}
{\dd P\over \dd x},
\eeq
\beq\label{hill2}
\left( {\dd \over \dd t} + \bb{w\cdot}\del\right){w_\phi}
+ 2\Omega w_R = - {1\over\rho} {\dd P\over \dd  y},
\eeq
\beq\label{hill3}
\left( {\dd \over \dd t} + \bb{w\cdot}\del\right){w_z}
= - z\Omega^2 - {1\over\rho } {\dd P\over \dd z}.
\eeq
(We have retained the vertical force in equation [\ref{hill3}].) 
Mass conservation is 
\beq\label{hill0}
{\dd\rho\over \dd t} + \del\bcdot(\rho\bb{w})= 0.
\eeq
In this form, equations (\ref{hill1}-\ref{hill0})
are sometimes referred to as the Hill equations.  

The Hill equations have a simple scaling symmetry:
\beq
\bb{w}(\bb{r}, t) \leftrightarrow\epsilon\bb{w}(\bb{r}/\epsilon,t),
\eeq
\beq
\rho (\bb{r}, t)
\leftrightarrow
\rho (\bb{r}/\epsilon, t),
\eeq
\beq
P (\bb{r}, t)
\leftrightarrow 
\epsilon^2 P (\bb{r}/\epsilon, t).
\eeq
In this form, with $\epsilon \ll 1$, we see that any solution of
the Hill equations that involves very small length scales has a
rescaled magnified counterpart with exactly the same time dependence.
Any instability at small scales would also have to be unstable at larger
scales.

\subsection{Shearing Coordinates}
We next introduce the coordinate transformation (Goldreich \& Lynden-Bell
1965):
\beq
\phi' = \phi -\Omega t = \phi -t[\Omega(R) + x {d\Omega/dR}],
\eeq
where we have expanded $\Omega$ in a Taylor series about radius
$R$ ($x$ is a small excursion from a fiducial value of $R$).
The other coordinate
transformations are identities, $t'=t$, $R'=R$, $z'=z$.
The transformed partial derivatives are
\beq
{\dd\ \over\dd t} +\Omega{\dd\ \over\dd \phi} = 
{\dd\ \over \dd t'},
\eeq
\beq
{\dd\ \over \dd R} = 
{\dd\ \over \dd R'}  - t {d\Omega\over dR}{\dd\ \over\dd\phi'},
\eeq
\beq
{\dd\ \over \dd z}=
{\dd\ \over \dd z'}, \qquad 
{\dd\ \over \dd \phi}=
{\dd\ \over \dd \phi'}.
\eeq
The explicit form of the equations of motion is
\beq\label{sh1}
{\dd u_R \over \dd t'} +
\bb{u\cdot}\del u_R - 2\Omega u_\phi = -{1\over\rho}\, 
\left(
{\dd\ \over \dd R'}  - t' {d\Omega\over dR}{\dd\ \over\dd\phi'}
\right)P,
\eeq
\beq\label{sh2}
{\dd u_\phi \over \dd t'} +
\bb{u\cdot}\del u_\phi + {\kappa^2\over 2\Omega} u_R = 
-{1\over\rho R} {\dd P\over\dd\phi'} ,
\eeq
\beq\label{sh3}
{\dd u_z \over \dd t'} +
\bb{u\cdot}\del u_z = - {1\over\rho} {\dd P\over \dd z'},
\eeq
where
\beq\label{sh4}
\bb{u\cdot}\del = u_R\left( {\dd\ \over \dd R'}
- t' {d\Omega\over dR}{\dd\ \over\dd\phi'}\right) + {u_\phi\over
R} {\dd\ \over\dd \phi'} + u_z{\dd\ \over\dd z'}.
\eeq
Mass conservation is
\beq\label{sh5}
\left(
{\dd\ \over \dd R'}  - t' {d\Omega\over dR}{\dd\ \over\dd\phi'}
\right) u_R + {1\over R}{\dd u_\phi\over\dd\phi'} +
{\dd u_z\over \dd z'} = 0.
\eeq
These equations form the basis of the analysis of the next section. 

\section {Analysis}

\subsection{Reduction to a Single ODE}

The system (\ref{sh1})--(\ref{sh5}) has no spatial dependence,
but does depend explicitly on time $t'$.  This suggests
that we seek a solution of the form
\beq\label{pw}
f(t')\exp\left[i\left(k'_R R' + m\phi'/R' + k_z z'\right)\right],
\eeq
where the function $f$ will of course differ for each of the velocity
components and the pressure.  Without loss of generality, we may
take $m>0$.  Notice that because of vanishing velocity divergence,
the nonlinear $\bb{(u\cdot\nabla) u}$ terms vanish explicitly.  Indeed,
the $\bb{u\cdot\nabla}$ operator acting upon any perturbed modal
plane wave vanishes.  One can immediately see why nonlinear rotational
hydro-instabilities are problematic:
linear theory and nonlinear theory
are the same for a finite amplitude plane wave.

Although a superposition of these shearing waves would not be a solution
of the nonlinear equations, in order to sustain turbulence such a
superposition would have to somehow access the free energy source of
differential rotation.  But in general, the neglected nonlinear 
$\bb{(u\cdot\nabla)u}$
wave term leads not to a viable free energy source, but to mutual
wave-wave interactions.  Absent a steady source term directly linked
to the differential rotation, it is perhaps not surprising that these
wave-wave cascades do not seem to be self-sustaining.


In what follows, the fluid velocity $\bb{u}$ and pressure $P$
variables refer to the time-dependent fourier $f$ amplitude of equation
(\ref{pw}).  We define the unprimed time-dependent wavenumber
\beq\label{k1}
k_R\equiv k_R (t') = k'_R -t'm{d\Omega\over dR},
\eeq
and shall henceforth drop the prime $'$ superscript from $t'$ and
other coordinates which are identical in both frames, and write
$d/dt$ for the time derivative of the amplitudes.  
The equations of motion are
\beq\label{k2}
{du_R\over dt} -2\Omega u_\phi =  - i k_R {P\over \rho},
\eeq
\beq\label{k3}
{du_\phi\over dt} + {\kappa^2\over 2\Omega} u_R = - i {m\over R} {P\over
\rho},
\eeq
\beq\label{k4}
{du_z\over dt} = - ik_z {P\over\rho},
\eeq
and the equation of mass conservation is
\beq\label{k5}
u_R k_R + m u_\phi/R + k_z u_z  = 0.
\eeq
An equation for $u_R$ can be extracted from equations (\ref{k1})--(\ref{k5})
after some straightforward, but tedious, algebra:
\beq
{d^2 u_R\over dt^2} - {4k_R m\over k^2 R} {d\Omega\over d\ln R}
{du_R\over dt} + 
\left[
{k_z^2\over k^2}\kappa^2 +  {2m^2\over k^2 R^2}
\left(d\Omega\over d\ln R\right)^2 \right]u_R = 0,
\eeq
where
\beq
k^2 \equiv k^2_R(t) + {m^2\over R^2} + k_z^2.
\eeq
This equation is in fact mathematically equivalent to equation (56)
of Johnson \& Gammie (2005a).  These authors have shown that it may
be solved in terms of hypergeometric functions.  Here, we prefer
to take a somewhat different tack, which avoids the introduction
of rather complicated special functions.
A pleasant simplification ensues if we introduce the
variable
\beq
U = k^2 u_R.
\eeq
Then, our differential equation takes on the compact Schr\"odinger
form:
\beq\label{ff}
{d^2U\over dt^2} + {k_z^2\over k^2} \kappa^2 U = 0,
\eeq
and is amenable to a classical WKB analysis.

\subsubsection{Explicit Solutions}

Equation (\ref{ff}) can be written in a convenient dimensionless form.
Let 
\beq
k_\perp^2 = k_z^2 + (m/R)^2,\quad k_\perp > 0 ,
\eeq
and introduce a new time variable $\tau$:
\beq
\tau \equiv {k_R(t)\over k_\perp}.
\eeq
Since $d\Omega/dR<0$ for the problems of interest, 
\beq
{d\tau\over dt} = - {m\over k_\perp}{d\Omega\over dR} > 0.
\eeq
With this substitution, our differential equation becomes
\beq\label{U}
{d^2U\over d\tau^2} + {\beta^2\over 1+\tau^2}U = 0,
\eeq
with 
\beq
\beta = \left|k_z\kappa/m\over d\Omega/d R\right|.
\eeq
The leading order WKB solution of this equation is
\beq
U \sim
\left(1+\tau^2\right)^{1/4}\, \exp\left[\pm i\beta\sinh^{-1}\tau\right],
\eeq
or
\beq\label{wkb}
u_R = u_R(0) \left(1+\tau^2\right)^{-3/4}\,
\exp\left[\pm\,  i\beta\sinh^{-1}\tau\right],
\eeq
where $u_R(0)$ is defined to be $u_R$ at $t=0$.  
Formally, this breaks down in the limit $\tau\gg \beta$, but in practice
very little changes.  The differential equation for $U$ may be easily
solved to leading order in the overlapping asymptotic domain $\tau \gg 1$.
For $\beta \ge 1/4$:
\beq\label{pl}
u_R =\mbox{constant}\times \tau^{-3/2} \exp\left[\pm\, i\beta(1-1/4\beta^2)^{1/2}\ln\tau
\right],
\eeq
while for $\beta \le 1/4$:
\beq\label{p1}
u_R =\mbox{constant}\times \tau^{-(3 \pm \sqrt{1-4\beta^2}\ )/2}.
\eeq
With the precise leading order
large $\tau$ solution in hand, we see that equation
(\ref{wkb}) is actually valid in the domain $\ln\tau\ll {8\beta}$.  For
$\beta=2$, for example, this corresponds to $\tau\ll 10^7$, considerably
larger than the naive domain of applicability, $\tau\ltsim\beta=2$.
Even beyond the extended domain, however, the WKB solution differs
from the actual large $\tau$ solution only by a slowly varying phase.
We shall henceforth adopt equation (\ref{wkb}) as our formal solution
for $u_R$.

\subsubsection
{Solution for the Pressure and Nonradial Velocities}

To complete the problem, we require expressions for
$u_\phi$, $u_z$, and $P$.  Unlike $u_R$, these flow variables need not
have a simple WKB form.  Nevertheless,
they may be obtained explicitly in terms of 
$u_R(\tau)$ and $du_R/d\tau$. 

The key is to note that equation (\ref{U})
may be written
\beq
u_R = -{1\over\beta^2} {d^2\ \over d\tau^2} \left[\left(1+\tau^2\right)u_R\right]
\eeq
which implies
\beq\label{uR}
\int u_R(\tau) \, d\tau = -{1\over\beta^2} {d\ \over d\tau}
\left[\left(1+\tau^2\right)u_R\right] + \mbox{constant.}
\eeq
If we combine equations (\ref{k3}), (\ref{k4}),  and
(\ref{k5}), there results
\beq\label{uphi}
{k_\perp^2\over k_z^2}{d u_\phi\over dt} + {\kappa^2\over 2\Omega}
u_R + {m\over Rk_z^2} {d(k_R u_R)\over dt} = 0.
\eeq
Writing this in terms of $\tau$ and simplifying,
\beq
{d\ \over d\tau}
\left(
{u_\phi } + {m \tau u_R\over k_\perp R} 
\right) +
{m\beta^2\over 2 k_\perp R} \left|d\ln\Omega\over d\ln R\right|
u_R =0.
\eeq
Finally, we integrate over $\tau$ and use equation (\ref{uR}).
We find
\beq
{u_\phi } + {m \tau u_R\over k_\perp R}\left( 1 - \left|d\ln\Omega\over d\ln R\right|
\right) - {m\over 2k_\perp R}\left|d\ln\Omega\over d\ln R \right|(1+\tau^2)
{du_R\over d\tau} = 0,
\eeq
where the integration constant on the right has been set to zero to be consistent
with the equation of motion (\ref{k2}).  This determines $u_\phi$ in terms
of $u_R$ and its first derivative:
\beq
u_\phi = {m\over k_\perp R} \left[
u_R\tau \left( \left|d\ln\Omega\over d\ln R\right|-1\right)+
{1+\tau^2\over 2}  \left|d\ln\Omega\over d\ln R\right|{du_R\over d\tau}
\right].
\eeq

Next, the axial velocity $u_z$ is obtained from $\del\bcdot\bb{u}=0$:
\beq
u_z = - {k_z\over k_\perp}
\left[
\tau u_R \left(1 + {m^2\over R^2k_z^2}\left|d\ln\Omega\over d\ln R\right|
\right)
+
{m^2\over 2 R^2k_z^2} \left|d\ln\Omega\over d\ln R\right|(1+\tau^2){du_R\over d\tau}
\right].  
\eeq
The enthalpy $P/\rho$ then follows from (\ref{k2}):
\beq
{P\over\rho} = -{2\Omega i \over k_\perp} 
\left[
u_R \left( \left|d\ln\Omega\over d\ln R\right|-1\right)
+
{m\tau \over 2 k_\perp R} \left|d\ln\Omega\over d\ln R\right| {du_R\over d\tau}
\right],
\eeq
where it is understood that the real part of the equation is to be taken.
This is convenient if $u_R$ is given as a WKB wave of the form (\ref{wkb}),
which is precisely how we shall use it. 
Choosing
\beq
u_R = {\rm Re}\left[{u_R(0)\exp[i\beta\sinh^{-1}(\tau)]\over (1+\tau^2)^{3/4}
}\right] = {u_R(0)\cos[\beta\sinh^{-1}(\tau)]\over (1+\tau^2)^{3/4}},
\eeq
we define the argument
\beq
X\equiv\beta\sinh^{-1}(\tau),
\eeq
and for Keplerian flow one
obtains
\beq
u_\phi = - u_R(0)\left({k_z\over k_\perp}
{\sin X\over2 (1+\tau^2)^{1/4}}
+{m\over k_\perp R}
{5\tau\cos X\over8(1+\tau^2)^{3/4}}
\right),
\eeq
\beq
u_z = u_R(0) 
\left[
{m\over2  k_\perp R} {\sin X\over (1+\tau^2)^{1/4}}
-{k_z\over k_\perp}
{\tau\cos X\over (1+\tau^2)^{3/4}}\left( 1 + {3m^2\over 8 k_\perp^2 R^2} \right)
\right],
\eeq
\beq
{P\over \rho} = {u_R(0)\Omega\over k_\perp}
\left[{k_z\over k_\perp}
{\cos X\over (1+\tau^2)^{5/4}}
+
{1+\tau^2(1-9m/4k_\perp R) \over (1+\tau^2)^{7/4}}\sin X 
\right].
\eeq
The velocity profiles are shown in Figure (1).

\begin {figure}
\plotone{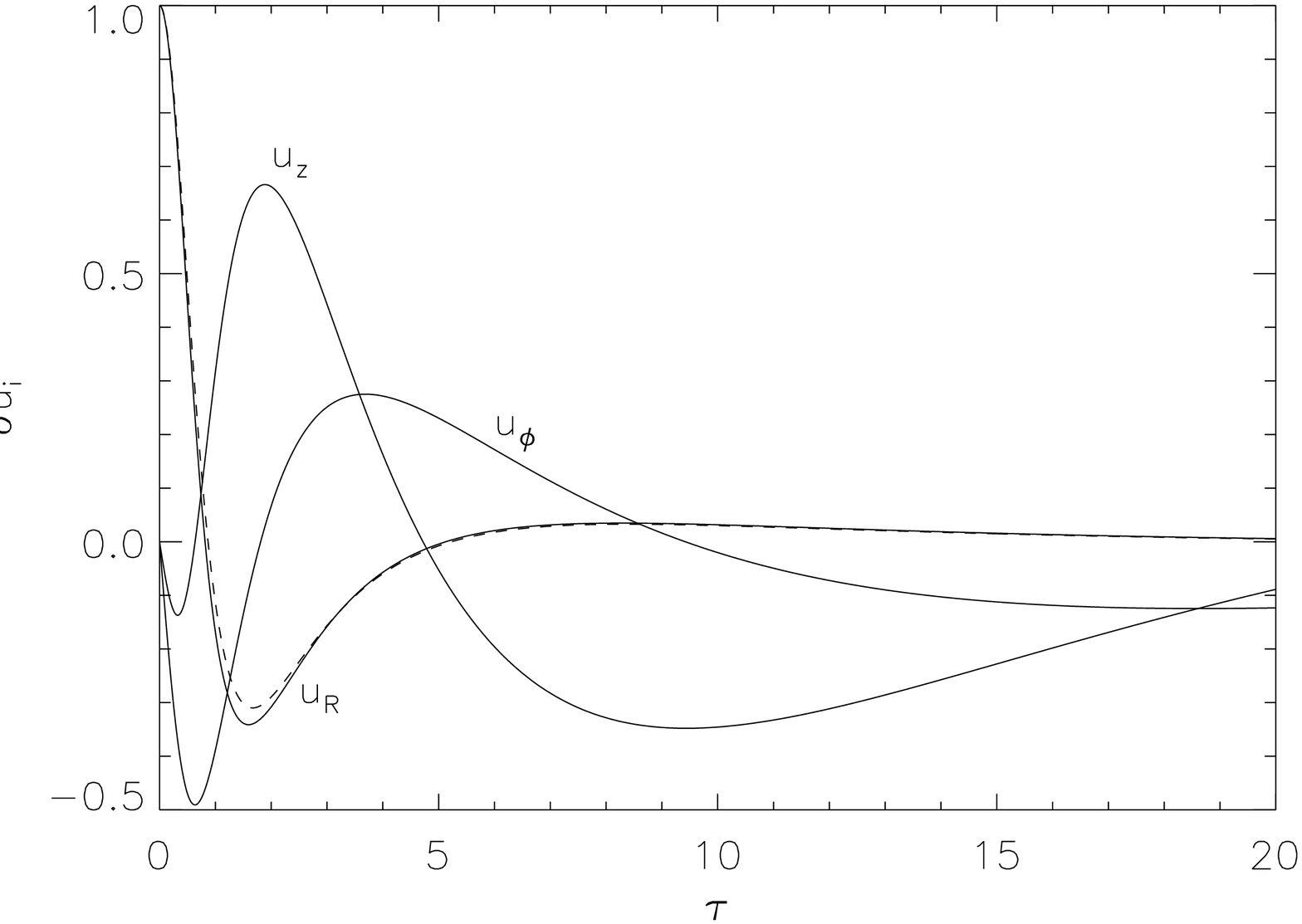}
\caption {Radial ($u_R$), azimuthal ($u_\phi$), and axial 
($u_z$) velocity components for disturbances 
in a $\beta=2$ Keplerian disk.  For reference, the (dashed line) 
radial velocity WKB solution lies
almost directly on top of the exact solution $u_R$, appearing
slightly above the true solution near the minimum point.}

\end{figure}

\subsection {The Small $\beta$ Limit}

The presence of the $\beta$ term is a strongly stabilizing influence
(it embodies epicyclic motion), and one therefore should examine the conditions
under
which it vanishes.  In a Keplerian disk $\kappa=\Omega$, and thus
$\beta\ll1$ must imply $k_zR\ll m$, which can correspond to one of
two different physical situations.
The first is when $u_z$ is a constant and all other variables
vanish.  These strictly vertical displacements are of little interest.
The second possibility corresponds to 
degenerate incompressible
two-dimensional disturbances
in the disk plane.  This problem has been considered by other authors
(Chagelishvili et al. 2003;  Johnson \& Gammie 2005a).
When $\beta \rightarrow 0$, we drop the second 
term in equation (\ref{U}), and find the formal result 
\beq
u_R \sim {C_1 +C_2\tau\over 1+ \tau^2},
\eeq
where $C_1$ and $C_2$ are constants of integration.   This is
consistent, of course, with equation (\ref{p1}) for large $\tau$.
But in this limit
equations (\ref{k2}), (\ref{k3}), and (\ref {k5}) may be directly combined
into the first order equation
\beq
{dU\over d\tau} = {d(k^2u_R)\over d\tau} = 0,
\eeq
so that $C_2$ must be zero.  (It represents a spurious solution.)
Chagelishvili et al. suggest that the growing phase of these solutions
($\tau<0$) may lead to turbulence, but we have seen that this solution
is in fact exact.  It cannot, and in simulations does not, lead to
sustainable turbulence.

All of this stands in stark contrast to the case of a vanishing specific
angular momentum gradient ($\kappa^2=0$), which does in fact lead to
turbulence.  Despite the apparently similar mathematics, the physical
behavior of this $\beta=0$ solution is very different.  Why?

Because in the case of $\kappa^2=0$, 
finite $\bb{k}=k_z\bb{e_z}$ 
{\em axisymmetric} modes now are algebraically unstable.
Without the steady march
toward arbitrarily large radial wavenumbers that cuts-off nonaxisymmetric
perturbations, axisymmetric $u_R$ perturbations continue to grow
linearly with time.   This solution is no longer
a spurious branch of equation (\ref{ff}), it is a valid solution
of the $m=\kappa=0$ limit of equations (\ref{k2})--(\ref{k5}).
(Note that $\beta$ is actually
undefined in this singular limit, and $\tau$ becomes
a constant; one must work directly with $t$.)
Perturbations in $u_\phi$ remain constant if
$m=0$, and grow linearly with time if the disturbance is even
the slightest bit nonaxisymmetric.
For a single mode, these are still exact nonlinear solutions, but
an ensemble of modes in this case does lead to a turbulent
cascade.  The difference here is that the cascade can be fed by the
axisymmetric instability, which features a consistently positive 
value for ${-u_Ru_\phi d\Omega/dR}$, and thus 
a robust source of free energy.
Indeed, in the axisymmetric limit, equations (\ref{k2}) and
(\ref{k4}) combine to give
\beq
u_R u_\phi = {1\over 4\Omega}{k^2\over k_z^2} {d u_R^2\over dt} > 0
\eeq
so that outward angular momentum transport accompanies the release of free
energy, growing linearly with time in the early stages.  By contrast,
non-axisymmetric solutions can maintain a source of free energy
only for {\em leading} waves; this source becomes a sink when the waves
become trailing.  Unlike the uniform
angular momentum turbulence, which is self-sustaining, fresh leading waves
are always needed in a Keplerian disks.  Without them,
perturbations die.

\section{Simulations}

Numerical simulations have sharpened our understanding of the stability
of Keplerian disks.  These calculations are not infallible, of course:
they are subject to the limitations that accompany any solution by
numerical approximation.  But the uncertainties due to numerical errors
can generally be minimized by resolution studies, a careful choice of
problem, and a detailed comparison with analytic solutions.  This is as
true for the hydrodynamic shearing box as for any other problem,
although there are not many known analtyic solutions that can serve as
tests.  The shearing wave solutions presented in this paper constitute
one such test problem.

Recently, the two-dimensional form of the shearing wave has been
simulated by Johnson \& Gammie (2005b), and by Shen, Stone \& Gardiner
(2006).  These two studies used two different simulation codes:
Johnson \& Gammie (2005b) used the operator-split, nonconservative
finite difference ZEUS algorithm (Stone \& Norman 1992), and Shen et
al. (2006) used the conservative Godunov Athena algorithm (Gardiner \&
Stone 2006).  Here, we have carried out a few simple simulations using
the Athena code to test the three-dimensional solutions developed in
this paper.  We have examined the evolution of a single linear shearing
wave using the piecewise parabolic (third-order) representation of the
underlying variables, and the Harten, Lax, van Leer with Contact (HLLC)
flux option of the Athena code (Gardiner \& Stone 2006).

In general, a wave propagated by a finite difference algorithm will be
subject to numerical truncation error, which is typically proportional
to some power of the ratio of the grid resolution divided by the
wavelength.  The Athena code has a formal truncation error that is
second order, and tests carried out by Shen et al. (2006) with
two-dimensional shearing waves showed convergence to the analytic
solution at a rate of $N^{-2.5}$ over a range from $N=64$ to 256 grid
zones.  An interesting aspect of this test problem is that the radial
wavenumber of a  shearing wave evolves with time as the wave becomes
increasingly tightly-wrapped (eq.~\ref{k1}).  Thus, the effective
resolution of any localized disturbance declines with time, regardless
of the number of grid zones used.  Shen et al. (2006) found that once a
wave passed to fewer than 16 grid zones per wavelength, its amplitude
decayed noticeably and smoothly as the resolution further diminished.
Since the dissipation in Athena is strongly nonlinear a wave will be
damped as its wavelength approached $2\Delta x$, where
$\Delta x$ is a fiducial grid interval; Shen et
al. observed no aliasing of trailing into leading waves.

The specific problem we have examined is a Keplerian shearing box with
$q=-3/2$, $\Omega = 0.001$, constant initial density $\rho =1$, using
both an isothermal equation of state with sound speed $c_s = 4.08
\Omega$, and an adiabatic equation of state with $\gamma = 5/3$ and
$P=10^{-6}$.  The computational domain has unit lengths, $L_R =L_\phi
=L_z = 1$.  In local Cartesian coordinates $dx=dR$,
$dy=Rd\phi$, the initial condition for the shearing wave is
\beq\label{initcon}
u_R = A_R \sin[2\pi(n_R x + n_\phi y + n_z z) ]
\eeq
where $n_i$ is the number of wavelengths within the domain: 
at $t=0$, $n_R =
0$, $n_\phi = 1$, and $n_z = 4$, corresponding to $\beta = 4$.  This
initial condition is particularly convenient:  $u_\phi$ and $u_z$ are
zero at $t=0$.  The initial wave amplitude $A_R$ is set to 
$10^{-5}$ for the adiabatic simulation.  We run the problem using
$64^3$, $128^3$, and $256^3$ grid zones.  Figure (2)
shows
the evolution of the three velocity components for this mode in the
$256^3$ simulation overlaid on the analytic solution (dashed line).
Some systematic deviation from the analytic solution becomes noticeable
beyond $n_R = 16$, which corresponds to 16 grid zones per radial
wavelength for the $256^3$ grid zone resolution.  The presence of
small amplitude, high frequency sound waves can also be seen.

\begin {figure}
\plotone{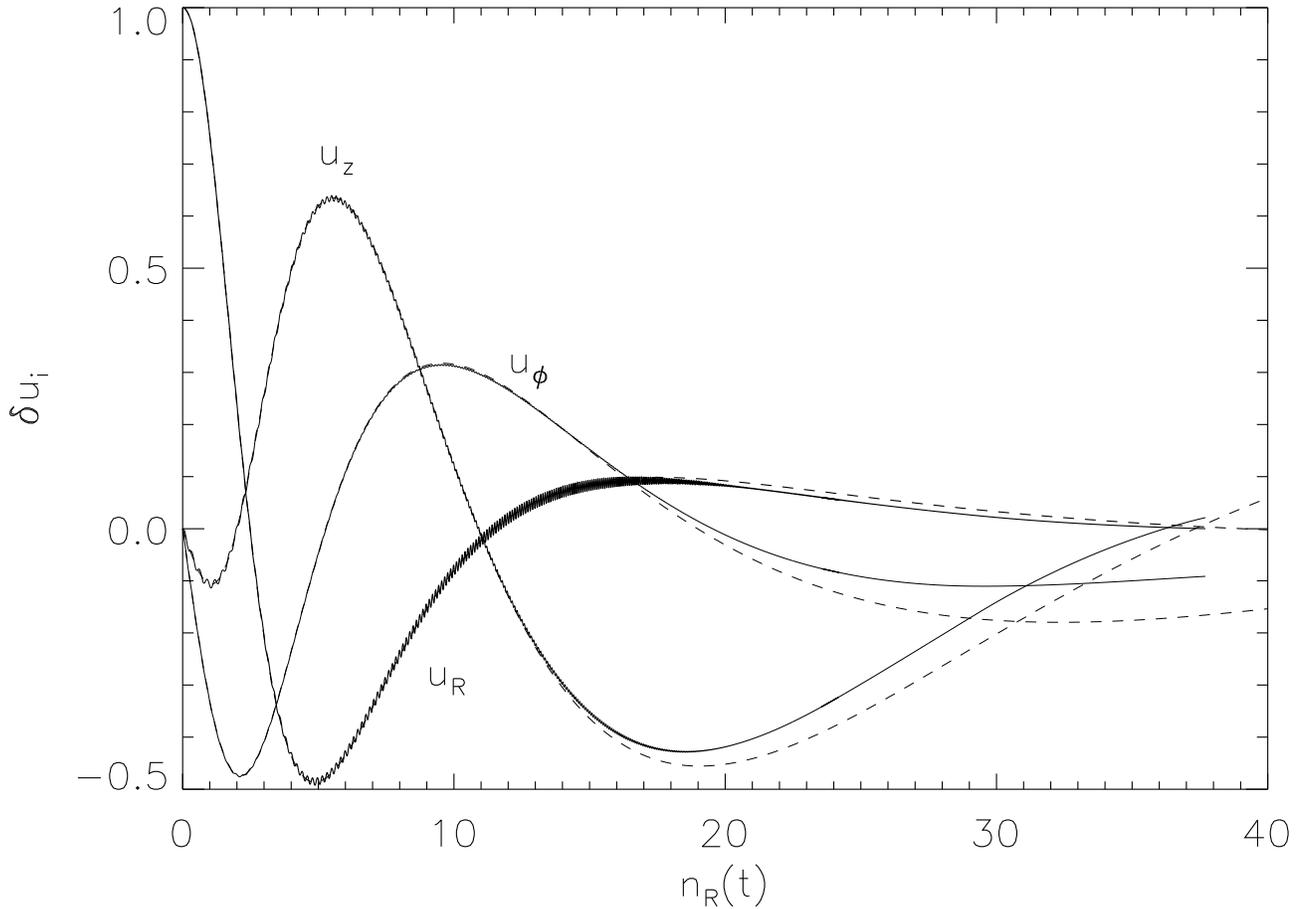}
\caption {Amplitudes for radial ($u_R$), azimuthal ($u_\phi$), and axial ($u_z$)
velocity components for a $\beta=4$ wave, computed in a Keplerian shearing
sheet using $256^3$ grid zones.  The dashed lines are the
exact solutions derived from direct integration of the ODE.
}
\end{figure}

The effect of resolution is illustrated in Figure (3)\ 
which plots the wave kinetic energy as a function of time for each of
the three grid resolutions along with the analytic value (dashed
line).  As one would expect, as resolution increases the numerical
solution maintains close adherence to the analytic value at higher
$n_R$ values.  Comparison of the fractional error for $n_R < 10$ shows
that between $128^3$ and $256^3$ zones the results are converging at
the expected second order rate.  The isothermal equation of state
simulations are similar, although they deviate from the analytic value
at an earlier time, suggesting that some of the difference is due to
compressibility effects in addition to numerical sources of error.

\begin {figure}
\plotone{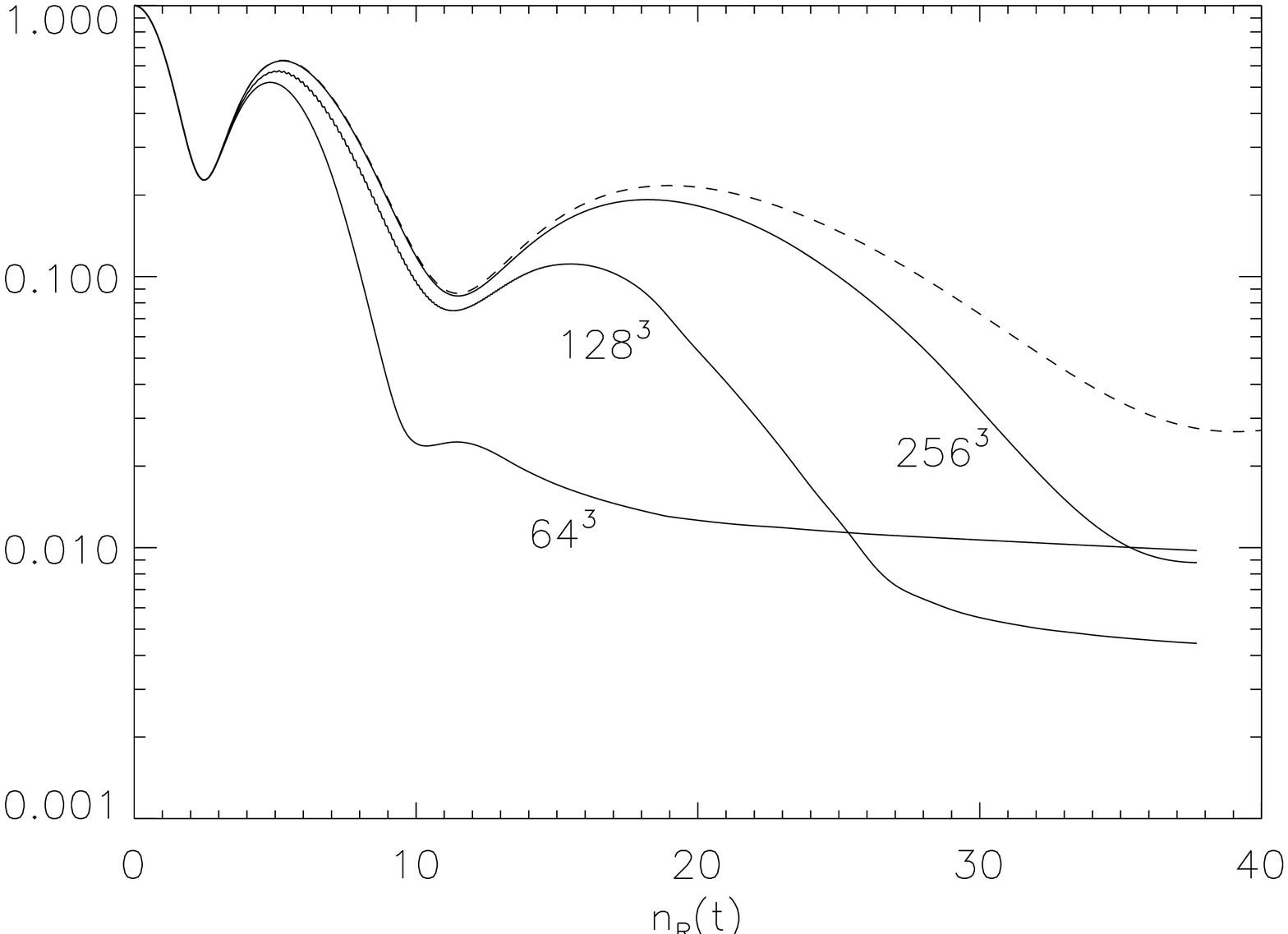}
\label{fig:energy}
\caption {Kinetic energy (as a fraction of the input value) in the
velocity perturbations for a $\beta=4$ Keplerian shearing sheet
simulation.  The labeled solid curves correspond to simulations using
$64^3$, $128^3$ and $256^3$ grid zones.  The dashed line is the
analytic solution for the kinetic energy.
}
\end{figure}

One reason that these shearing waves were of interest in previous
two-dimensional studies was that they are amplified during their 
leading phase.  They reach a peak amplitude when $\tau = 0$; beyond
that point they are trailing, and their amplitude declines.  Numerical
simulations offer a way to investigate the effects of compressibility
and the stability of the waves as their peak velocity amplitudes
approach the level of the background shear or the sound speed.  Using
$1024^2$ grid zone two-dimensional simulations, Shen et al. (2006)
examined large amplitude waves and found that, for either leading or
trailing waves, the waves  become unstable when
\begin{equation}
\label{ssgeqn}
|k_R  u_\phi | \gtsim |d\Omega/d\ln R|.
\end{equation}
When this limit is exceeded the waves rapidly break down.

As a simple study of large amplitudes in three dimensions we start the
$\beta = 4$ shearing wave mode at $\tau = 0$ as above, but with a large
amplitude, namely $A_R = 0.002$.  At this amplitude the stability
criterion (\ref{ssgeqn}) will be violated when $\delta u_\phi$ reaches
its peak at $n_R \sim 2$.  This wave was evolved using $64^3$, $128^3$
and $256^3$ grid zones.  Figure (4) 
shows the evolution of the kinetic energy in the velocity perturbations
as a function of time.  All three resolutions show a breakdown of the
wave solution, although with increasing resolution the energy tracks
the analytic solution for a slightly longer period of time before
breaking down.  The wave undergoes what appears to be a Kelvin-Helmholtz
instability (Shen et al. 2006).  The radial and azimuthal velocity
amplitudes drop rapidly, leaving mainly vertical motions that have a
finite $n_R$ but are nearly axisymmetric and $z$-independent.  This flow
subsequently breaks down, again in a manner resembling a Kelvin-Helmholtz
instability, and the wave kinetic energy rapid decays.  Interestingly,
the higher resolution simulation shows the most rapid decline in perturbed
kinetic energy.  Thus, while the high resolution simulations
track the analytic solutions longer, they breakdown faster.  This is
perhaps a reflection of the fact that the highest resolved wavenumbers
have the most rapid Kelvin-Helmholtz growth rates.  The lower resolution simulations
take longer for the secondary instability to affect the vertical columns.
As with the two-dimensional simulations, at late times the perturbed
kinetic energy drops toward zero as does the (positive) Reynolds stress.
There is no hint of sustained turbulence.

\begin {figure}
\plotone{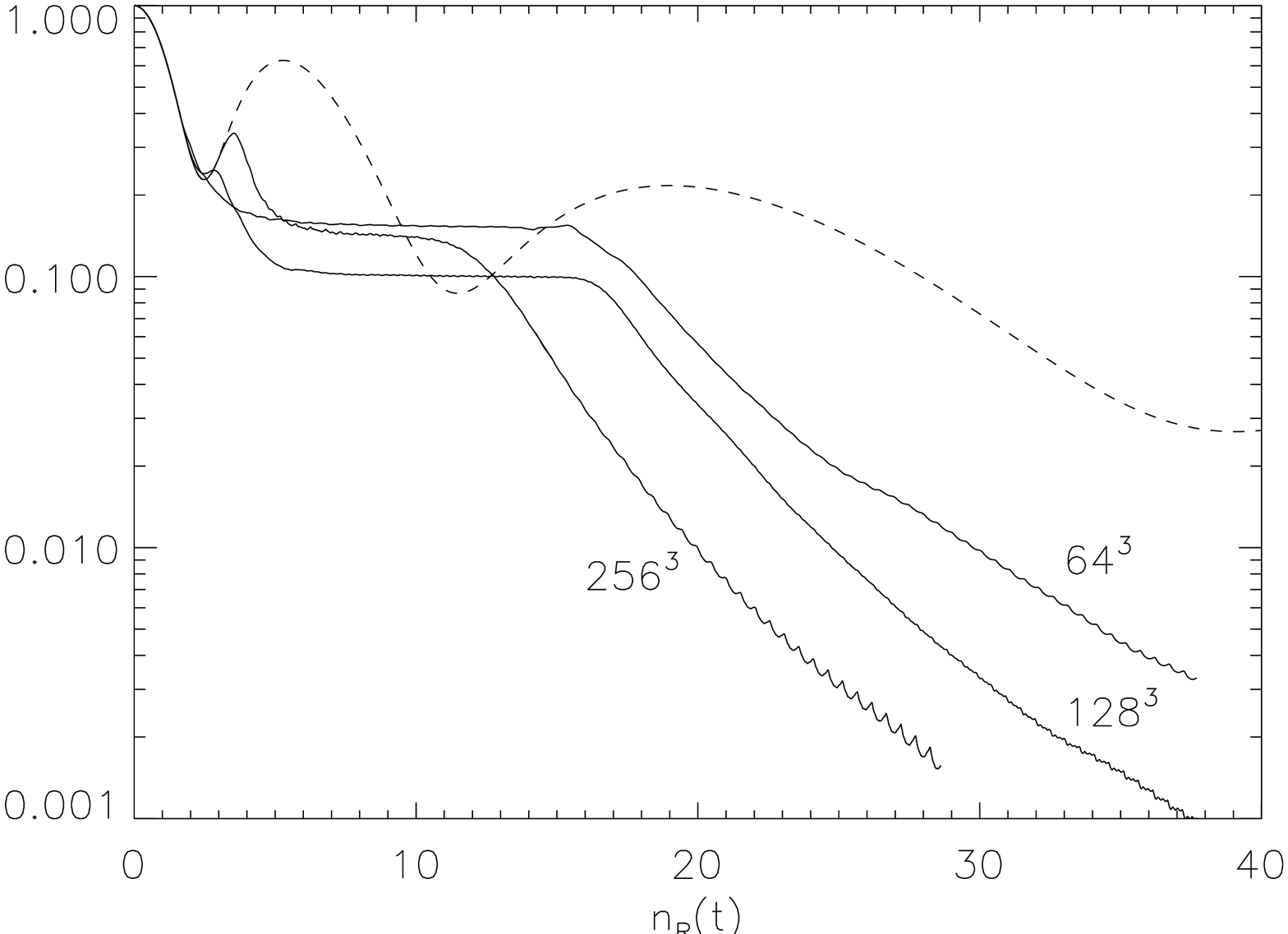}
\label{fig:decay}
\caption {Kinetic energy (as a fraction of the input value) in the
velocity perturbations for a $\beta=4$ Keplerian shearing sheet
simulation that began with a large initial amplitude.  Shown are curves
for $64^3$, $128^3$ and $256^3$ grid zones.  The dashed line is the
analytic solution for the kinetic energy.
}
\end{figure}

\section {Discussion and Conclusion}

We have argued that the growing behavior of the $\tau<0$ phase of local
disk perturbations that has been put forth as a possible origin of
sustained Keplerian turbulence will not, in fact, lead to such behavior.
It is true that an arbitrarily large leading wavenumber will formally
produce an arbitrarily large amplification factor during this phase
of the evolution.  The difficulty is that the linear and nonlinear
behavior of the equations is the same for these incompressible plane wave
solutions.  The fact that the amplitudes become temporarily large does not
change their wave character.  Moreover, it appears that in the initial
strongly leading configuration, the large shear in the wave is unstable,
dissipating the disturbance at small amplitude before it has time to grow.

True disk turbulence requires a tap into a permanent source of free energy,
not large wave amplitudes.  It is misleading to
view the MRI and shear layer turbulence as a consequence of
large amplitude perturbations.  These flows are turbulent
because there is a free energy path between the differential rotation
and the perturbations, and it is already present at small
amplitudes.  The ensuing turbulent cascade is generally not be characterized
by large amplitude disturbances.

The outcome of numerical simulations form a coherent and interrelated
pattern of results.  For example, Stone \& Balbus (1996) studied the
behavior of convective turbulence in Keplerian disks as a source of
angular momentum transport.  The correlation tensor $\langle u_R
u_\phi\rangle$, a measure of the angular momentum flux, was found
to be tiny.  Moreover, it had the wrong sign (radially inwards).
Though not a direct investigation of Keplerian stability, these
results make sense only if the underlying disk is stably stratified,
and numerical diffusion is largely absent:  reverse angular momentum
transport is just what is expected in any Rayleigh-stable nonmagnetized
disk that is driven externally (Balbus 2001).  Detailed direct studies
of Keplerian stability by Balbus et al. (1996), and Hawley et al. (1999)
demonstrated nonlinear stability once more, even when the disks were
strongly nonlinearly driven.  Proponents of transient amplification as
a route to turbulence must argue that the very large and tightly wound
wavenumbers needed to enter into the nonlinear regime are inaccessible
to current codes.   But if {\em this} were the problem, one could
simply start the calculation in the nonlinear regime.  Flows that are
nonlinearly unstable are in fact quite sensitive to starting amplitudes.
When this is done for Keplerian disks, no transitions to sustained
turbulence are found.  Indeed, excellent convergence of all stable flows
was found in the study by Hawley et al. (1999), even though the codes
involved had completely different dissipative properties.


The fact that the plane wave solution we have found is exact to all
orders in its amplitude suggests that coupling between the disturbance
and the background is limited to simple wave action conservation.
This is not a route to turbulence.  A superposition of such plane wave
solutions would exhibit group velocity dispersive spreading and nonlinear
wave-wave interactions; the latter would impart power to smaller scales.
Unless the fluid is itself externally driven, however, there is nothing
to sustain the fluctuations against these losses.

Because our solution is inviscid and three-dimensional, it provides an
excellent diagnostic for probing the large dynamical range properties of
numerical schemes.   The solution is strictly exact only for constant
density fluids, but the velocity components of an adiabatic fluid are
accurately described by our solution.  The large $|\tau|$ behavior and
peak amplitudes are stringent tests of a numerical scheme.  The {\em
disagreement} between the analytic and numerical large $|\tau|$ behavior
is particularly revealing, since it shows that the codes are able to
uncover more complicated secondary features, even at very small scales.
The most obvious effect in this work is compressibility, but the small
scale Kelvin-Helmholtz instability reported by Shen et al. (2006)
during the tightly-wrapped phase of the wave evolution (when $|u_R|
\ll |u_\phi|$) was unforeseen.  A strongly leading shearing wave never
even reaches the predicted analytic amplitude: its initial intrinsic
shear is too unstable.

Numerical, analytic, and laboratory studies appear to have reached a
robust consensus: absent additional flow dynamics, Keplerian rotation
is locally nonlinear stable.

\section*{Acknowledgements}

This work was supported in part by NASA grants NAG5--13288 and
NNG04--GK77G.   SAB also received support from a Chaire d'Excellence
award from the Ministry of Higher Education of France.   We thank Tom
Gardiner and James Stone for valuable discussions on both the Athena
code and shearing waves, and the referee for constructive comments.
The simulations were carried out on the TeraGrid system at NCSA.

\end{document}